\documentclass[prl,twocolumn,showpacs,superscriptaddress,preprintnumbers,amsmath,amssymb,amsfonts]{revtex4}
\usepackage{graphicx}% Include figure files
\usepackage{dcolumn}% Align table columns on decimal point
\usepackage{bm}% bold math
\usepackage{times}

\begin{document}

%\preprint{APS/123-QED}

\title{Spike synchronization of a chaotic array as a phase transition}
\author{M. Ciszak}
\affiliation{C.N.R.-Istituto Nazionale di Ottica Applicata, L.go
E. Fermi 6, 50125, Firenze, Italy}
\author{A. Montina}
\affiliation{Dipartimento di Fisica, Universit\`{a} di Firenze,
Via Sansone 1, 50019 Sesto Fiorentino (FI), Italy}
\author{F. T. Arecchi}
\affiliation{C.N.R.-Istituto Nazionale di Ottica Applicata, L.go
E. Fermi 6, 50125, Firenze, Italy} \affiliation{Dipartimento di
Fisica, Universit\`{a} di Firenze, Via Sansone 1, 50019 Sesto
Fiorentino (FI), Italy}

\date{\today}

\begin{abstract}
We study how a coupled array of spiking chaotic systems
synchronizes to an external driving in a short time.
Synchronization means spike separation at adjacent sites much
shorter than the average inter-spike interval; a local lack of
synchronization is called a defect. The system displays sudden
spontaneous defect disappearance at a critical coupling strength.
Below critical coupling, the system reaches order at a definite
amplitude of an external input; this order persists for a fixed
time slot. Thus, the array behaves as an excitable system, even
though the single element lacks such a property. The above
features provide a dynamical explanation of feature binding in
perceptual tasks.
\end{abstract}

%Use showkeys class option if keyword
\maketitle

%\section{\label{intro} Introduction}

Temporal versus rate coding for the neural-based information has
been open to debate in the neuroscience literature~\cite{1}. The
electrical activity of a single neuron is measured by
micro-electrodes inserted in the cortical tissue of
animals~\cite{2,2a}. This activity consists of trains of action
potentials or "spikes". In rate coding only the mean frequency of
spikes over a time interval matters, thus  requiring a suitable
counting interval, which seems unfit for  fast decision tasks.
Temporal coding assigns importance to the precise timing and
co-ordination of spikes. A special type of temporal coding is
synchrony, whereby information is encoded by the synchronous
firing of spikes of selected neurons in a cortical module.

Let an animal be exposed to a visual field containing two separate
objects. Since each receptive field isolates a specific detail,
one should expect a corresponding large set of different
responses. On the contrary, all the cortical neurons whose
receptive fields are pointing to the same object synchronize their
spikes, and as a consequence the visual cortex organizes into
separate neuron groups oscillating on distinct spike trains for
different objects ({\it feature binding})~\cite{2}. Indirect
evidence of synchronization has been reached for human subjects as
well, by processing the EEG (electro-encephalo-gram)
data~\cite{4}.

Since feature binding results from the readjustment of the
temporal positions of the spikes, a plausible explanation is based
on the mutual synchronization properties of chaotic oscillators;
in fact, an erratic distribution of the spike occurrence prior to
an applied stimulus seems mandatory for adapting to different
rhythms (\cite{5}, and references therein).

In this Letter we show that spiking chaotic dynamics provides a
suitable model of the feature binding process, where fast
synchronization is necessary for completion of a perceptual
task~\cite{2,4}. Indeed, the conscious perception of a feature
(e.g. form or motion) is believed to be due to the synchronization
of the neurons of a cortical module within a limited time
slot~\cite{zeki}; the time limitation is necessary to leave room
to successive perceptions. Feature binding implies two temporal
aspects~\cite{2,4,5}, namely, i) it is associated with gamma band
EEG oscillations ($25$ msec mean interspike separation), and ii)
it lasts a few hundred milliseconds. Since we intend to model the
neurons of a cortical module during a perceptual task, we study
the transient dynamics of defects in the coupled system. A defect
is defined as the lack of synchronization of two adjacent sites.
Defect disappearance marks the global synchronization of a
modulus. The two main results of this paper are: i) above a
critical coupling a chaotic array fully synchronizes within tens
of interspike intervals; ii) in presence of an input, the array
displays a collective recognition which lasts for a fixed time.

We focus on a model exhibiting homoclinic chaos (HC)~\cite{7} and
compare it with  a generic chaotic system as e.g.
R\"{o}ssler~\cite{6}. Let us consider an array of HC systems with
nearest neighbor bidirectional coupling ruled by
\begin{eqnarray}
\dot{x}_1^i&=&k_0x_1^i(x_2^i-1-k_1\sin ^2x_6^i) \nonumber\\
\dot{x}_2^i&=&-\gamma _1 x_2^i-2k_0x_1^ix_2^i+gx_3^i+x_4^i+p \nonumber \\
\dot{x}_3^i&=&-\gamma _1x_3^i+gx_2^i+x_5^i+p\nonumber \\
\dot{x}_4^i&=&-\gamma _2x_4^i+zx_2^i+gx_5^i+zp \label{eq1} \\
\dot{x}_5^i&=&-\gamma _2x_5^i+zx_3^i+gx_4^i+zp\nonumber \\
\dot{x}_6^i&=&-\beta
\left[x_6^i-b_0+r\left(f(x_1^i)+\epsilon(x_1^{i-1}+x_1^{i+1}-
2\eta ^i(t))\right)\right] \nonumber
\end{eqnarray}
where $f(x_1^i)=\frac{x_1^i}{1+\alpha x_1^i}$ and $\eta ^i(t)$ is
a variable obeying the filter equation $\dot{\eta }^i=-d(\eta
^i-x_1^i)$ with $d=10^{-3}$. The index $i$ denotes the $i$th site
position for $i=1,...,M$. The values of parameters are:
$k_0=28.5714$, $k_1=4.5556$, $\gamma _1=10.0643$, $\gamma
_2=1.0643$, $g=0.05$, $p=0.016$, $z=10$, $\beta =0.4286$, $\alpha
=32.8767$, $r=160$ and $b_0=0.1032$. Their physical meaning has
been discussed in Ref.~\cite{7}. This parameter choice is by no
means critical, since several successive finite chaotic windows
are available. The mutual coupling consists of adding to the $x_6$
equation on each site a function of the intensity $x_1$ (action
potential) of the neighboring oscillators. Chaos due to the
homoclinic return to a saddle focus implies a high sensitivity to
an external perturbation in the neighborhood of the
saddle~\cite{5}.

\begin{figure}[h]
\begin{center}
%\vspace{-0.5cm}
\includegraphics*[width=1.\columnwidth]{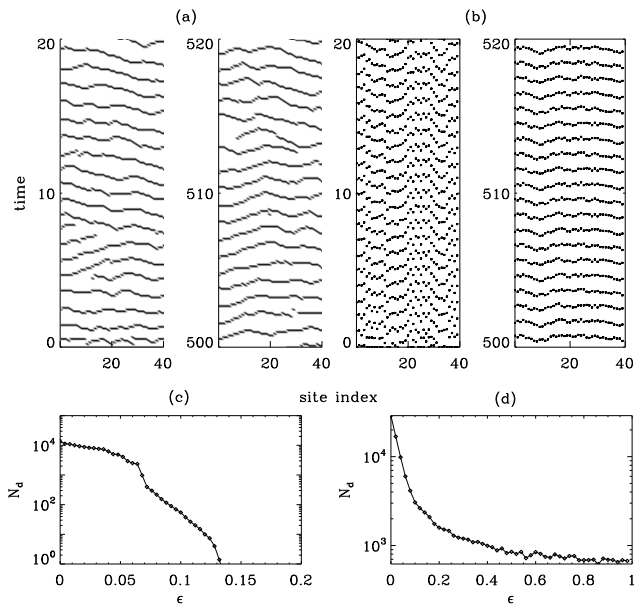}
\vspace{-0.8cm} \caption{\label{fig1} Space-time plots of (a) HC
systems for $\epsilon =0.1$ where time is in $\langle ISI\rangle $
units and (b) R\"{o}ssler systems for $\epsilon =0.01$ where time
is in $T$ units. The degree of synchronization is characterized by
the number of defects $N_d$ estimated far beyond the initial
transient: (c) $100$ coupled HC; (d) $40$ coupled R\"{o}ssler.}
\end{center}
\end{figure}
At each pseudo-period, or inter spike interval = $ISI$, HC yields
the alternation of a regular large spike and a small chaotic
background. The chaotic background is the sensitive region where
the activation from the neighbors occurs, while the spike provides
a suitable signal to activate the coupling. We use the generic
attribution of HC for a large class of systems, with a saddle
focus instability~\cite{8}. This includes a class B laser with
feedback~\cite{7}, a modified Hodgkin-Huxley model for action
potentials in a neuron membrane~\cite{13} and the Hindmarsh-Rose
model of generation of spike bursts~\cite{14}. In view of this
high sensitivity, we expect that they synchronize not only under
an external driving~\cite{9}, but also for a convenient mutual
coupling strength. A quantitative indicator of this sensitivity is
represented by the so-called propensity to
synchronization~\cite{10}. Furthermore, for an array of coupled
systems, the onset of a collective synchronization was explored
numerically for different values of the coupling strength
$\epsilon $, showing that the lack of synchronization manifests
itself as phase slips, that is one spike less or more compared to
the adjacent site over the same time interval. In a space-time
plot, space denoting the site position and time the point-like
occurrence of spikes, phase slips appear as
dislocations~\cite{11}. Space-wise, a tiny variation of $\epsilon
$ changes dramatically the size of the synchronized
domain~\cite{11}. Here we explore the dynamical mechanisms
underlying the onset of synchronization, both the spontaneous one,
occurring in a coupled array above a critical $\epsilon $ in the
absence of an external input, as well as the stimulated one
(semantic response) induced by an external stimulus localized at
one site. Stimulated array synchronization occurs for couplings
below $\epsilon _c$.
\begin{figure}[h]
\begin{center}
\includegraphics*[width=1.\columnwidth]{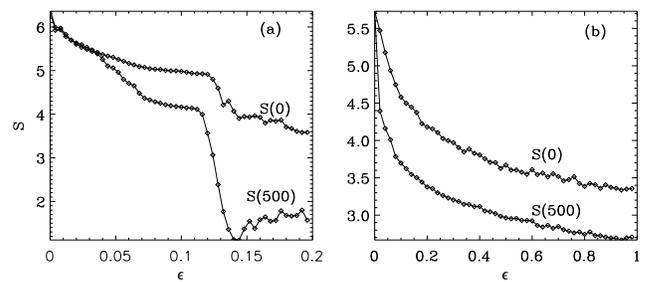}
\vspace{-0.7cm} \caption{\label{fig2} The degree of
synchronization of a coupled array estimated during the initial
transient characterized by entropies $S(0)$ and $S(500)$ versus
$\epsilon $ for (a) $100$ coupled HC and (b) $40$ coupled
R\"{o}ssler.}
\end{center}
\end{figure}

We analyze the interaction of neurons by taking the large
amplitude spikes as $1$ and the remaining chaotic and refractory
background as $0$. Then we define the response time $t_r$ as the
time difference between spike occurrence at neighboring sites
starting from the first. When neurons are uncoupled, their spikes
are uncorrelated and the distribution of the response times
(including defects) is spread over a broad range. Increasing the
coupling strength correlated clusters appear, being however
characterized by a large variance in the response times and
boundary appearance of defects. Indeed, a large fluctuation in the
response time corresponds to a defect which interrupts a sequence
of synchronized sites, representing the boundary with between two
clusters. Increasing further the coupling strength we observe the
transition to overall synchronization, accompanied by equal
response times for all sites.  The synchronization is not
isochronous, as the response time distribution consists of two
symmetric non zero peaks. This is due to a time lag starting from
the two end sites, which obey open boundary conditions. In the
case of unidirectional coupling we still have the synchronization
without isochronism ~\cite{17}, but this time with a single $t_r$
peak.

Spiking systems appear to be the operating units of the brain
cortex~\cite{1}. In order to find out what is the peculiarity of
spiking systems in the synchronization processes, we compare the
spiking dynamics of HC with another chaotic system, namely
R\"{o}ssler. The oscillations in R\"{o}ssler are characterized by
a leading single frequency $1/T$ and chaos appears in the
amplitude of oscillations. Here, no spike but rather phase
synchronization occurs. This implies a time code with a poorer
resolution, as a spike duration in HC is less than one tenth the
$\langle ISI \rangle $, whereas phase is resolved over a sizable
fraction of the period $T$. The dynamics is ruled by
\begin{eqnarray}
\dot{x}_1^i&=&-x_2^i-x_3^i+\epsilon (x_1^{i-1}+x_1^{i+1}-2x_1^i)\nonumber \\
\dot{x}_2^i&=&x_1^i+ax_2^i \nonumber \\
\dot{x}_3^i&=&b+x_3^i(x_1^i-c)\label{eq3}
\end{eqnarray}
where $a=0.15$, $b=0.2$ and $c=10$. We analyze the R\"{o}ssler
systems in a similar manner: when the oscillation crosses the zero
line upward we take it as a localized $1$ (spike-like); otherwise
it is $0$. Also in this case a non-isochronous synchronization can
be observed, however for strong couplings the system tend to
synchronize isochronously.
\begin{figure}[h]
\begin{center}
\includegraphics*[width=0.65\columnwidth]{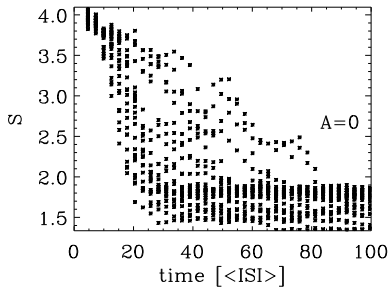}
\vspace{-0.6cm} \caption{\label{fig4} Entropy $S$ for $100$
coupled HC calculated throughout the initial time slot, starting
from different initial conditions, in the absence of external
driving and for $\epsilon =0.136$.}
\end{center}
\end{figure}

We are interested in the way the defects decay in HC and
R\"{o}ssler. In Fig.~\ref{fig1} we report the space-time series
starting from $t=0$ as well as from $t=500$ (time being in
$\langle ISI \rangle $ or $T$ units); space is the linear site
sequence. For HC with $\epsilon=0.1$ in both panels of
Fig.~\ref{fig1}a we observe the same number of defects on average,
thus there is no decay of defects. We obtain the same results also
for longer times (not shown here). Defects die out at one site and
are born again at another one in course of time, due to the
temporal relations between adjacent sites necessary to induce the
escape from the saddle region (see the detailed discussion for
unidirectional coupling in Ref.~\cite{17}). In R\"{o}ssler the
appearance of defects is a global phenomenon: the non zero phase
difference between neighboring sites lasts after a long time (left
panel in Fig.~\ref{fig1}b). Once the array establishes the
synchronization line, defects do not reappear again as seen in
right panel of Fig.~\ref{fig1}b.

We characterize the degree of synchronization in HC and
R\"{o}ssler in terms of two quantities, namely, the number of
defects $N_d$ in a space-time interval corresponding respectively
to the $40$ sites and $20\langle ISI \rangle$ and the entropy $S$.
As the control parameter we take the coupling strength $\epsilon
$. In HC, $N_d$ decreases with $\epsilon$ and goes to zero at the
critical coupling strength $\epsilon _c=0.13$ (Fig.~\ref{fig1}c).
In R\"{o}ssler instead, defects decay monotonically with the
coupling strength (Fig.~\ref{fig1}d) but no sharp transition is
observed. The degree of order of the system is obtained by
calculation of the entropy $S$ from the distributions of the
response times $t_r$ in the time series during or far beyond the
initial transient. The entropy $S$ is defined as
\begin{eqnarray}
S(t)=-\sum_{t_r}p(t_r,t)\ln p(t_r,t) \label{eq2}
\end{eqnarray}
where $p(t_r,t)$ is the normalized probability distribution of
$t_r$ evaluated at time $t$. As done for $N_d$, we take for
$p(t_r,t)$ the average over intervals of $20\langle ISI \rangle$,
starting at $t$.

In the case of HC (Fig.~\ref{fig2}a) $S$ has a sharp transition at
$\epsilon = \epsilon _c$. The transition appears exactly at the
critical coupling for which the number of defects goes to zero
(Fig.~\ref{fig1}c). The transition to synchronization during the
transient time has similar characteristics as the synchronization
far beyond the transient (see Fig.~\ref{fig2}a). Also in this case
the sudden decrease of entropy at the critical point is associated
with the total disappearance of defects. When the coupling
strength is set above $\epsilon _c$, defects decay immediately
after the coupling is switched on.
\begin{figure}[h]
\begin{center}
\includegraphics*[width=1\columnwidth]{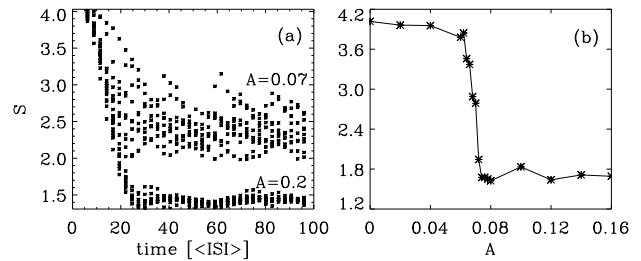}
\vspace{-0.7cm} \caption{\label{fig5} HC array, $\epsilon=0.11$:
(a) Entropy vs time for different amplitudes of an external signal
with period $T=0.5\langle ISI\rangle$; (b) long time entropy vs
input amplitude $A$, for the same $T$.}
\end{center}
\end{figure}
Indeed, below $\epsilon _c$ the occurrence of defects hinders
synchronization between sites. The reason is that after the
appearance of a defect, the system needs some time to establish
synchronization again. Such appearance and disappearance of
defects (or equivalently, synchronization and de-synchronization
of spikes) is responsible for the lack of synchronization in the
system at small coupling strengths. Finally, we checked that the
value of the critical coupling is universal for any size of the
array. The reason for that is the locality of the bidirectional
coupling here considered.

Now we check if the transition occurs also in R\"{o}ssler. We
estimate the entropy $S$ versus $\epsilon $ during the initial
transient and far beyond it, as in the case of HC. In
Fig.~\ref{fig2}b the decrease of entropy $S$ versus $\epsilon$ is
smooth, without sudden changes. It is because defects decay
steadily during the time evolution. In this case, once the
synchronization is established, defects do not reappear as it
happens instead in HC. On the other hand, as shown in
Fig.~\ref{fig2}b, defects appear during the transient time up to
large values of the coupling strength.

%\section{Defects dynamics and phase transition}
Thus, there exists a crucial difference between spike and phase
synchronization, namely, in HC defects appear and disappear
constantly on average up to $\epsilon _c=0.13$. Beyond $\epsilon
_c$ defects never reappear. On the contrary, in R\"{o}ssler, there
are no critical changes in the defect dynamics, and for non-zero
coupling strengths they always decay in the course of the time
evolution. The defect dynamics in HC appears as a new feature; in
fact in sudden symmetry changes of space extended Ginzburg-Landau
type systems transient defects decay always, as shown
theoretically~\cite{18} and verified experimentally in
He$^3$~\cite{19} and in nonlinear optics~\cite{20}. The difference
in defect dynamics in HC and R\"{o}ssler highlights the existence
of the phase transition in the synchronization of the HC systems.

%\section{Time-dependent transition to synchronization}
The initial transient is interesting from a neurological point of
view, since perception implies synchronization within the short
time slot as remarked above. We must then explore a {\em transient
phase transition}, requiring that all sites be synchronized within
a limited time following the application of an external signal.
\begin{figure}[ht]
\begin{center}
\includegraphics*[width=1.\columnwidth]{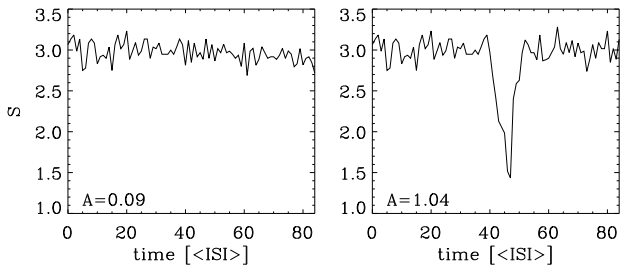}
\vspace{-0.6cm} \caption{\label{fig6} The entropy in time for HC
array perturbed at only one site by a pulse with amplitude
$A=0.09$ (left panel) and $A=0.104$ (right panel). In both cases
duration of pulse is $\Delta t=\langle ISI\rangle $ and the
coupling strength $\epsilon =0.104$.}
\end{center}
\end{figure}
We set the coupling strength slightly above the critical value
$\epsilon _c=0.13$ and measure the entropy $S$ of the initial time
slots ($t=0$) for different sets of initial conditions. The
results are reported in Fig.~\ref{fig4}. There are two things to
be noticed. First, the way in which the transients differ: the
spread of entropies has a maximum some time after the coupling has
been switched on and before the transition to synchronization
appears. This deviation is due to the different competition
histories between the coupled sites. Second, the existence of
different final values of entropy after the transition to
synchronization. This difference is determined by the existence of
different ordered states (attractors)~\cite{20a}.

From Fig.~\ref{fig4} it appears that the ordered states are
reached after $20\langle ISI\rangle$. Since feature binding has
been related to the gamma band neuronal oscillations, with
$\langle ISI\rangle\sim25 ms$ ~\cite{2,4}, and a perceptual task
requires a few hundred milliseconds~\cite{4}, such a task has to
be over within less than $30\langle ISI\rangle$.

In order to model the perceptual task, the transition to
synchronization has to be accomplished in the presence of the
external driving. Let us consider a periodic signal applied at the
first site of the array. We introduce the signal by modulation of
parameter $b_0$ as: $b=b_0(1+A\sin 2\pi t/T )$. Entropy depends on
the four parameters $\epsilon$, $t$, $A$ and $T$. Leaving to
Ref.~\cite{20a} a detailed study, here we comment on the following
issues. In Fig.~\ref{fig5}a, we plot the time behavior of $S$ for
$\epsilon=0.11$, below spontaneous synchronization, and show that
an input with period $T=0.5\langle ISI\rangle$ yields a
synchronized array for $A=0.2$, whereas, for lower $A$, the
interval $20$ to $80\langle ISI\rangle$ is still at high entropy.
In Fig.~\ref{fig5}b, we fix $t=20\langle ISI\rangle$,
$\epsilon=0.11$, $T=0.5\langle ISI\rangle$ and plot the dependence
of $S$ on $A$. A sharp discontinuity occurs around
$A_c\simeq0.07$. Here a novel feature emerges: while each
individual system is chaotic and slightly perturbed by the input,
the global array behaves as an excitable system, with a sharp
transition from a steady high value of entropy to a low one. At
variance with previous models which consist of arrays of excitable
individuals, usually taken as FitzHugh-Nagumo
systems~\cite{tsuda}, here the excitable behaviour emerges as a
collective property of the coupled array. To test the excitable
property of an array, we show in Fig.\ref{fig6} how stimuli of
amplitude $A$ (above $A_c$) and duration $\Delta t $ induce an
excitable collective response. The response has a form of
synchronization lasting a finite time; then the array comes back
to its previous state.
%Conclusions

In summary, we have demonstrated a phase transition in  coupled
chaotically spiking systems (HC) and characterized this transition
in terms of the disappearance of defects at the critical coupling
strength $\epsilon _c$. Below $\epsilon_c$ and in the absence of a
stimulus, defects persist in course of time. On the other hand in
chaotic non-spiking systems as R\"{o}ssler, where synchronization
regards the phase, and not the spike positions, defects decay at
long times even for very small coupling strengths. The important
feature of the spiking chaotic systems as HC is that transition to
synchronization in response to an external driving is sudden, as
it occurs in every day experiences related to recalling the memory
of events or objects (we recall a record in a step manner: either
we record it or not). We attribute this sudden coherent response
to a global excitability property.

This research was supported by a Marie Curie Intra-European
Fellowships within the 6th European Community Framework Programme
and by Ente Cassa di Risparmio di Firenze under the Project
Dinamiche cerebrali caotiche.


\begin{thebibliography}{99}
\bibitem{20a} The details regarding this problem will be reported elsewhere.
\end{thebibliography}
\end{document}